\documentclass[aps, twocolumn, showpacs, amsmath,superscriptaddress]{revtex4}
\usepackage{amssymb}
\usepackage{dcolumn}
\usepackage{bm}
\usepackage{graphicx}
\usepackage{color}
\usepackage{ulem}
\DeclareGraphicsExtensions{. jpg,. pdf, . mps, . png, .eps, . ps, . EPS}

\begin{document}
\def\be{\begin{equation}}
\def\ee{\end{equation}}

\def\bc{\begin{center}} 
\def\ec{\end{center}}
\def\bea{\begin{eqnarray}}
\def\eea{\end{eqnarray}}
\newcommand{\avg}[1]{\langle{#1}\rangle}
\newcommand{\ket}[1]{\left |{#1}\right \rangle}
\newcommand{\beq}{\begin{equation}}
\newcommand{\eneq}{\end{equation}}
\newcommand{\beqnn}{\begin{equation*}}
\newcommand{\eneqnn}{\end{equation*}}
\newcommand{\beqy}{\begin{eqnarray}}
\newcommand{\eneqy}{\end{eqnarray}}
\newcommand{\beqynn}{\begin{eqnarray*}}
\newcommand{\eneqynn}{\end{eqnarray*}}
\newcommand{\half}{\mbox{$\textstyle \frac{1}{2}$}}
\newcommand{\proj}[1]{\ket{#1}\bra{#1}}
\newcommand{\av}[1]{\langle #1\rangle}
\newcommand{\braket}[2]{\langle #1 | #2\rangle}
\newcommand{\bra}[1]{\langle #1 | }
\newcommand{\Avg}[1]{\left\langle{#1}\right\rangle}
\newcommand{\inprod}[2]{\braket{#1}{#2}}
\newcommand{\upket}{\ket{\uparrow}}
\newcommand{\downket}{\ket{\downarrow}}
\newcommand{\Tr}{\mathrm{Tr}}
\newcommand{\hcontrol}{*!<0em,.025em>-=-{\Diamond}}
\newcommand{\hctrl}[1]{\hcontrol \qwx[#1] \qw}
\newenvironment{proof}[1][Proof]{\noindent\textbf{#1.} }{\ \rule{0.5em}{0.5em}}
\newtheorem{mytheorem}{Theorem}
\newtheorem{mylemma}{Lemma}
\newtheorem{mycorollary}{Corollary}
\newtheorem{myproposition}{Proposition}
\newcommand{\vp}{\vec{p}}
\newcommand{\Or}{\mathcal{O}}
\newcommand{\so}[1]{{\ignore{#1}}}

\newcommand{\red}[1]{\textcolor{red}{#1}}
\newcommand{\blue}[1]{\textcolor{blue}{#1}}

\title{Supersymmetric multiplex networks described by coupled Bose and Fermi statistics}

\author{Ginestra Bianconi}

\affiliation{School of Mathematical Sciences, Queen Mary University of London, London E1 4NS, United Kingdom}

\begin{abstract}
Until now,   no simple symmetries have been detected  in complex networks.
Here we show that, in growing multiplex networks  the symmetries of  multilayer structures can be  exploited  by their dynamical  rules, forming supersymmetric multiplex networks described by coupled Bose-Einstein  and Fermi-Dirac quantum statistics. 
The supersymmetric multiplex is formed  by layers which are  scale-free networks and can display a Bose-Einstein condensation of the links.
To characterize the complexity  of the supersymmetric multiplex using quantum information tools, we extend the definition of the network entanglement entropy to  the layers of multiplex networks.  Interestingly we observe a very simple relation between the entanglement entropies of the layers of the  supersymmetric multiplex network and the entropy rate of the same multiplex network. This relation therefore connects the classical non equilibrium growing dynamics of the supersymmetric multiplex network with its quantum information  static characteristics. 
\end{abstract}


\maketitle
\section{Introduction}
Recently, the relation among complex networks,  their geometry  and evolution \cite{RMP,Newman_rev,Boccaletti2006},  and   more traditional fields of physics  such as  quantum physics and quantum communication \cite{Cirac,Calsamiglia,Sachdev,Ising,Hubbard,JCH}, quantum information \cite{Quantuminformation}, cosmology \cite{Cosmology} and quantum gravity \cite{Rovelli,Rovelli2,triangulations} are starting to attract the interest of scientists.  

For instance, an interesting aspect of complex network evolution is that in major examples of realistic models for complex networks growth, 
quantum statistics emerge as important distributions determining the network dynamics.
In fact, the  evolution of scale-free networks with fitness of the nodes \cite{Fitness} is described by the  Bose-Einstein statistics and this model can be therefore mapped to a Bose gas. In correspondence of the  Bose-Einstein condensation \cite{Bose} of the Bose gas, the network topology undergoes a major structural condensation transition in which a node is linked to a finite fraction of links. This model can explain the emergence of ``super-hubs" in complex networks and the  ``winner-takes-all" phenomenon.  Interestingly  the evolution of  Cayley trees with fitness  of the nodes, is determined by the Fermi-Dirac statistics \cite{Fermi,Complex}. In addition to that, it has also been found that ensembles of complex networks  have been shown to be related with quantum statistics \cite{Garlaschelli}.  Finally, it has been shown that random geometric networks on a hyperbolic space can show a scale-free network topology \cite{Hyperbolic,Navigability} and that growing networks in hyperbolic space define the  so called ``network cosmology" \cite{Cosmology}. 
 From the quantum information perspective, complex networks encode relevant information in their structures, and new quantum entropy measures have been proposed in order to characterize and quantify this information \cite{Vonneuman,Garnerone,Jesus2}.Moreover, the  quantum random walk can be used to propose  different definitions of quantum PageRank on networks \cite{Garneronegoogle,Jesus1,Faccin,Zimboras,Caldarelli}.

Multiplex networks are multilayer network structures that are attracting large interest \cite{PhysicsReports,Kivela}. In fact they are able to describe several  types of interactions between the same set of nodes. For example, social networks, where the same people are connected by different means of communications, or different types of social ties, like friendship, collaboration, co-authorship, and so on are better described by multiplex networks.
Similarly,  if we want to describe diffusion in transportation networks we need to consider the multiplex nature of the underlying transportation networks, where  a given  location can be linked to other locations by different types of transportation, train, bus, airplane etc.
Finally, in brain networks the large variety of neuron types and types of interactions between them, will   not be fully understood  if the multilayer approach is not adopted \cite{Bullmore,Makse,Liaisons}.
Multilayer have a highly non trivial structure including communities \cite{Mason} and many different types of encoded structural correlations \cite{PRE,Goh,Vito}. Recently  some quantum information measures such as the Von Neumann entropy of single networks \cite{Vonneuman} have been extended to multilayer networks in order to quantify their complexity  \cite{Math}.

Multiplex networks are formed by layers that are usually scale-free and have a number of nodes that increases in time. For these reasons their evolution can be described by growing multiplex networks models \cite{Growth1,Growth2,Nonlinear}.
In single layer networks growing network models \cite{BA,Fitness,Bose} are able to explain the spontaneous emergence of scale-free degree distribution. In particular   the most fundamental growing network model, the Barab\'asi-Albert (BA) model \cite{BA} includes  only two dynamical rules: growth and preferential attachment, meaning that at each time a new node is added to the network and it attaches links preferentially to high degree nodes.
Just these two elements of the model have been shown to be responsible for a scale-free degree distribution of exponent $\gamma=3$. 
An important element determining network evolution is the intrinsic quality of the nodes, determining their fitness that make them more likely to acquire new links, with respect to other nodes with the same degree \cite{Fitness}.
The dynamics of single  growing scale-free networks with fitness of the nodes can be mapped to a Bose gas \cite{Bose}  giving rise to the intriguing phenomenon of the Bose-Einstein condensation, while growing Cayley trees with fitness of the nodes  can be mapped to a Fermi gas\cite{Fermi} by using a similar mathematical formalism.

In the context of multiplex networks, growing multiplex networks models have been shown   to generate multiplex networks with different degree distributions in the different layers, and different pattern of correlations \cite{PRE, Goh,Vito} between the degrees of the same node in different layers \cite{Growth1,Growth2,Nonlinear}. Nevertheless the role of the fitness of the nodes on growing multiplex networks models has not yet been explored.
Here we combine the process of link addition and the process of rewiring of the links in presence of an intrinsic fitness of the nodes. We show that the network growth can  exploit the symmetries of the multiplex networks and we can  generate scale-free supersymmetric multiplex networks described by coupled Fermi-Dirac and Bose-Einstein statistics. Moreover, we explore  the information content of these structures with quantum information tools by extending the definition of entanglement entropy of single networks \cite{Garnerone} to multiplex structures. To this end we map each each network in a given layer to a quantum network state and we  describe the complexity of each layer of the  multiplex networks by calculating the entanglement entropies of these quantum states. Finally  we relate  the entanglement entropies  of the layers of the supersymmetric multiplex network to  the entropy rate of the multiplex network.
The entropy rate \cite{Entropyrate} of the supersymmetric multiplex model  describes the rate at which the typical number of multiplex networks, which can be realized during the dynamics, grows in time.
Therefore here we show  that the entanglement entropies of the layers of the supersymmetric network, which  extract information  form a snapshot of the multiplex network using quantum information tools,  have a very simple relation with the entropy rate of the same supersymmetric  multiplex network, describing the classical non-equilibrium dynamics of the multilayer structure. 

The paper is structured as follows:
In section II we define the supersymmetric multiplex model and we provide its mean-field solution.
In section III we characterize the supersymmetric multiplex structural properties.
In section IV we define the entropy rate of the supersymmetric multiplex network.
In section V evaluate the  entanglement entropies of the layers  of the supersymmetric multiplex network.
In section VI we relate the entanglement entropies of the layers of the supersymmetric multiplex network  with its entropy rate.
Finally in section VII we give the conclusions.

We note here that for simplicity in  the main text of the paper  we consider only the case of a duplex (i.e. a multiplex with $M=2$), but the analysis can be easily extended to a multiplex with a generic value of the number of layers $M$.
Therefore in  the appendix the extension to supersymmetric multiplex networks with generic value of $M$ is discussed.
\section{ supersymmetric multiplex network model}
\subsection{The supersymmetric multiplex evolution}
A multiplex network is a multilayer system  formed by $N$ nodes having a copy (or replica) in each of the  $M$ layers, and $M$ layers formed by different networks of interactions between the $N$ nodes.
Here we assume that each node $i$ has quenched quality quantified by the  quenched parameter $\epsilon_i\in (0,1)$ called the energy of the node and we indicate by 
\bea
\eta_i=e^{-\beta\epsilon_i}
\eea
the fitness of the node $i$, determined both by its energy and a global parameter $\beta>0$.
On single layers it has been shown \cite{Fitness,Bose} that the fitness of the nodes is able to explain how some nodes (the more fit) acquire links at a faster rate than others, as observed in a variety of complex networks, including the Internet and the World-Wide-Web.
In particular, in these models it is assumed that  each new node links preferentially to nodes of high degree and high fitness, with a probability $\Pi_i$ that it attaches a new link to a node $i$ given by  
\bea
\Pi_i=\frac{\eta_i k_i}{\sum_j \eta_jk_j}=\frac{e^{-\beta\epsilon_i}k_i}{\sum_j e^{-\beta\epsilon_j}k_j},
\eea
where $k_i$ is the degree of node $i$, in this way generalizing the dynamics of the famous BA model \cite{BA} in which the preferential attachment is only driven by the degree of the nodes.

When considering the evolution of technological, social or transportation multilayer networks, a similar mechanism can be taken into account.
Nevertheless t,he new links of  a given layer might be added also according to a preferential attachment mechanism deriving from the popularity of a node in another layer. This model the case in which   a popular node in one layer  attracts also links in another layer.  Moreover, it can also happen that, instead, the popularity of a node in a layer is modulated by a process of rewiring of the links, and that links attached to popular nodes in one layer are more likely to be rewired. This could mimic the case in which the connectivities of hubs are damped by the process of  link rewiring.
We consider for simplicity a multiplex networks formed by $M=2$ layers and we describe its evolution as a growing multiplex network model in which each node has a given energy and corresponding fitness.  Specifically, we consider the case in which in the dynamics in the two networks is not symmetrical. In one layer new links are exclusively added according to a generalized preferential attachment which rewards high fitness nodes of high degree in either one of the layers, and  in the other layer the links are attached but also rewired in such a way to reduce the growth of the degree of high fitness nodes in that layer. This model can be generalized to a multiplex networks with larger number of layers in which the layer can be divided in two groups, each group of layer behaving in a similar way. In order to keep the description of the model simple we now focus on the case in which $M=2$, discussing in the appendix  the generalization to the general value of $M$. 
We   start at $t=0$ from a small set of nodes $N_0$ connected in both layers. 
Each node $i$ of the network has degrees $k^{[1]}_i,k^{[2]}_i$ respectively in layer 1 and layer 2, and energy $\epsilon_i$ drawn from a $g(\epsilon)$ distribution.
At each time  $t$ we add a node  to the multiplex network, each node has two replicas nodes, one on each layer. Moreover, each replica node is  initially attached to $m$ existing nodes in the same layer. In the following we will  indicate with $l$ the number of network changes, i.e. links additions or link rewirings,  occurring in each layer  starting from time $t=1$. After time $t$ we will have $l=mt$.
For each   network change in layer 1  we follow the subsequent procedure:
\begin{itemize}
\item We extract a number $\sigma^{[1]}_l=1,2$. The event $\sigma^{[1]}_l=1$ occurs  with  probability  $p^{[1]}(1)=\alpha$ while the event $\sigma_l^{[1]}=2$ occurs with probability   $p^{[1]}(2)=1-\alpha$.
\item If $\sigma^{[1]}_l=1$   the new node is attached  in layer 1  to a node $i^{[1]}_l$ chosen with probability
\bea
\Pi^{[1,1]}\left(i_l^{[1]}=i\right)=\frac{e^{-\beta \epsilon_i}k_i^{[1]}}{\sum_j e^{-\beta \epsilon_j}k_j^{[1]}},
\eea
i.e. it will be attached preferentially to nodes with low energy and high degree in layer 1,  according to a generalized preferential attachment.
Instead, if $\sigma^{[1]}_l=2$ the new  node  is attached in layer 1 to a node $i^{[1]}_l$ chosen with probability
 \bea
\Pi^{[1,2]}\left(i_l^{[1]}=i\right)=\frac{e^{-\beta \epsilon_i}k_i^{[2]}}{\sum_j e^{-\beta \epsilon_j}k_j^{[2]}},
\eea
i.e. will be attached preferentially to nodes with low energy and high degree in layer 2,  according to  a generalized preferential attachment.
\end{itemize}
For each  network change in layer 2 we follow the subsequent procedure:
\begin{itemize}
\item  We extract a number $\sigma^{[2]}_l=1,2$.
 The event $\sigma^{[2]}_l=1$ occurs  with  probability  $p^{[2]}(1)=1-\alpha$ while the event $\sigma_l^{[2]}=2$ occurs with probability   $p^{[2]}(2)=\alpha$. 
\item If $\sigma^{[2]}_l=1$  the new node will be attached, in  layer 2, to a node $i_l^{[2]}$ chosen with probability
\bea
\Pi^{[2,1]}\left(i_l^{[2]}=i\right)=\frac{e^{-\beta \epsilon_i}k_i^{[1]}}{\sum_j e^{-\beta \epsilon_j}k_j^{[1]}}.
\eea
Instead, if $\sigma^{[2]}_l=2$ a random link of a node $i^{[2]}_l$ chosen with probability 
\bea
\Pi^{[2,2]}\left(i_l^{[2]}=i\right)=\frac{e^{-\beta \epsilon_i}k_i^{[2]}}{\sum_j e^{-\beta \epsilon_j}k_j^{[2]}}
\eea
 is rewired, i.e. it is detached from node $i_l^{[2]}$ and attached to the new node of the network.
\end{itemize}
Therefore the network is  determined by the sequence of the values $\{\sigma^{[1]}_l,i_l^{[1]}\sigma^{[2]}_l,i_l^{[2]}\}$ that fully determines the evolution of the multiplex network given the initial condition.

\subsection{Mean-field solution of   the supersymmetric multiplex  model}
When studying growing networks with preferential attachment, in general large attention is given to the degree sequence of the network.
In order to predict the degree distribution of these models, mean-field approaches have been extensively studied, finding that in general they give a very good prediction of the structural properties  of the network \cite{Doro_book}.
In this paper we analyse the supersymmetric multiplex model  with the mean-field theory  leaving to subsequent works the analysis with the master equation approach. In order to check the validity of the approach we then compare the analytical results to simulations as discussed in Section \ref{sim}.
In the mean-field approach, one assumes that the degree of each node has no fluctuations, and therefore identifies the degrees $k^{[1]}_i,k^{[2]}_i$ at time $t$ with their average over the multiplex network realization. Moreover this approximation is also called the continuous approximation because it is assumed that both the degrees of the nodes and the time are continuous variables.
Therefore the mean-field equations for the supersymmetric multiplex model read
\bea
\frac{dk_i^{[1]}}{dt}=me^{-\beta \epsilon_i}\left[\frac{\alpha}{\sum_j e^{-\beta\epsilon_j}k_j^{[1]}}k_i^{[1]}+\frac{(1-\alpha)}{\sum_j e^{-\beta\epsilon_j}k_j^{[2]}}k_i^{[2]}\right],\nonumber \\
\frac{dk_i^{[2]}}{dt}=me^{-\beta \epsilon_i}\left[\frac{(1-\alpha)}{\sum_j e^{-\beta\epsilon_j}k_j^{[1]}}k_i^{[1]}-\frac{\alpha}{\sum_j e^{-\beta\epsilon_j}k_j^{[2]}}k_i^{[2]}\right].
\label{mf0}
\eea
Using an approach similar to the one used in the Bianconi-Barab\'asi model \cite{Bose}, we will assume self consistently that
\bea
\lim_{t\to \infty}\frac{\sum_j e^{-\beta\epsilon_j}k_j^{[1]}}{mt}&=&c_1, \nonumber \\
\lim_{t\to \infty}\frac{\sum_j e^{-\beta\epsilon_j}k_j^{[2]}}{mt}&=&c_2,
\label{selfc0}
\eea
where $c_1$ and $c_2$ are constants independent of the network realization.
Therefore, asymptotically in time we have 
\bea
\sum_j e^{-\beta\epsilon_j}k_j^{[1]}&\simeq& mtc_1,\nonumber \\
\sum_j e^{-\beta\epsilon_j}k_j^{[2]}& \simeq &mtc_2.
\label{asymptotics}
\eea

If we define the vector of the degree of each node as 
\bea
{\bf k}_i=\left(\begin{array}{c} k^{[1]}_i(t)\nonumber \\ k^{[2]}_i(t)\end{array}\right),
\eea
and we substitute for $t\gg1$ the asymptotic expression for the normalization sums Eq. $(\ref{asymptotics})$ we obtain that the mean-field Eqs. $(\ref{mf0})$ can be written as 

\bea
\frac{d {\bf k}_i}{dt}=\frac{e^{-\beta\epsilon_i}}{t}{\bf A} {\bf k}_i,
\label{mfs}
\eea
where the matrix ${\bf A}$ is defined as
\bea
{\bf A}&=&\left(\begin{array}{cc}\alpha/c_1 & (1-\alpha)/c_2\nonumber\\ (1-\alpha)/c_1&-\alpha/c_2\end{array}\right).
\eea
The solution of Eq. $(\ref{mfs})$ is given in terms of the eigenvalue and the eigenvector of the matrix ${\bf A}$.These eigenvalue are respectively  positive and  negative for every value of the parameter $\alpha$ of the model.
 We will indicate the eigenvalues of ${\bf A}$ as   $\lambda_{+}$ and $\lambda_{-}$ in correspondence of their sign.
These eigenvalues are given by 
\bea
\lambda_{+}=\frac{1}{2}\left[\left(\frac{\alpha}{c_1}-\frac{\alpha}{c_2}\right)+ \sqrt{\Delta}\right],\nonumber \\
\lambda_{-}=\frac{1}{2}\left[\left(\frac{\alpha}{c_1}-\frac{\alpha}{c_2}\right)- \sqrt{\Delta}\right],
\eea
with 
\bea
\Delta=\left(\frac{\alpha}{c_1}-\frac{\alpha}{c_2}\right)^2+4\frac{\alpha^2+(1-\alpha)^2}{c_1c_2}.
\eea
We have therefore that the constants $c_1$ and $c_2$ can be expressed as a function of $\lambda_{+}$ and $\lambda_-$ as
\bea
c_1&=&\frac{(1-\alpha)^2+\alpha^2}{2\alpha}\left[\left(\frac{1}{\lambda_+}+\frac{1}{\lambda_-}\right)+\sqrt{\Lambda}\right],\nonumber \\
c_2&=&\frac{(1-\alpha)^2+\alpha^2}{2\alpha}\left[-\left(\frac{1}{\lambda_+}+\frac{1}{\lambda_-}\right)+\sqrt{\Lambda}\right],
\eea
with 
\bea
\Lambda&=&\left(\frac{1}{\lambda_+}+\frac{1}{\lambda_-}\right)^2-\frac{4\alpha^2}{[(1-\alpha)^2+\alpha^2]}\frac{1}{\lambda_+\lambda_-}.
\eea
Moreover,  we  indicate by ${\bf u^+}=(u_1^{+},u_2^{+})$ and ${\bf u^-}=(u_1^-,u_2^-)$ the eigenvectors corresponding respectively to the eigenvalues $\lambda_{+}$ and $\lambda_-$.
The components of these eigenvectors are given by 
\bea
u^{+}_{1}&=&\frac{c_1}{2(1-\alpha)}\left(\frac{\alpha}{c_2}+\frac{\alpha}{c_1}+\sqrt{\Delta}\right), \nonumber \\
u^{+}_{2}&=&1,\nonumber \\
u^{-}_1&=&\frac{c_1}{2(1-\alpha)}\left(\frac{\alpha}{c_2}+\frac{\alpha}{c_1}-\sqrt{\Delta}\right),\nonumber \\
u^{-}_2&=&1.
\eea
Therefore, solving the Eqs. $(\ref{mfs})$ the  degrees of node $i$ in the two layers can be calculated in the mean-field approximation to be 
\bea
{\bf k}_i(t)=d^+ {\bf u^{+}}\left(\frac{t}{t_i}\right)^{e^{-\beta \epsilon_i}\lambda_{+}}+d^{-} {\bf u^{-}}\left(\frac{t}{t_i}\right)^{-e^{-\beta\epsilon_i}\lambda_{-}},
\label{k1}
\eea
where $t_i$ is the time at which the node $i$ is arrived in the network and 
where $d^-$ and $d^+$ are constants determined by the initial condition ${\bf k}_i(t_i)=m{\bf 1}$ where ${\bf 1}$ is the column vector of components $(1,1)$. Starting from Eq. $(\ref{k1})$, the initial condition can be also written as 
\bea
{\bf k}_i(t_i)=m{\bf 1}={\bf U}{\bf d},
\eea
where ${\bf U}$ is the matrix with column vectors given by the eigenvectors ${\bf u}^{+}$ and ${\bf u}^{-}$ i.e. 

\bea
{\bf U}=\left(\begin{array}{cc}u_{1}^{+}& u_{1}^{-}\nonumber \\
u_{2}^{+}&u_{2}^{-}\end{array}\right),
\eea
and the column vector ${\bf d}$ has components ${\bf d}=(d^{+},d^{-})$.
This equation can  always be solved finding that the constants $d^{+}$ and $d^{-}$ are given by 
\bea
{\bf d}&=&m{\bf U}^{-1}{\bf 1}\nonumber \\
&=&\frac{m}{u_1^{+}u_2^{-}-u_1^{-}u_2^{+}}\left(\begin{array}{c}u_2^{-}-u_1^{-} \nonumber \\ u_1^{+}-u_2^{+}\end{array}\right).
\label{dp}
\eea
Note that the denominator of Eq. $(\ref{dp})$ is always positive definite and never singular  since we have $u_2^{-}>0,u_2^{+}>0,u_1^{+}>0$ but $u_1^{-}<0$.
Having fixed the constant $d^{+}$ and $d^{-}$,  
we can rewrite Eq.~$(\ref{k1})$ as 
\bea
{\bf k}_i={\bf B}{\bf v}_i
\label{k2}
\eea
with ${\bf v}_i$ indicating the column vector 
\bea
{\bf v}_i=\left(\begin{array}{c} \left(\frac{t}{t_i}\right)^{e^{-\beta \epsilon_i}\lambda_+}\\\left(\frac{t}{t_i}\right)^{e^{-\beta \epsilon_i}\lambda_-}\end{array}\right),
\eea
and the matrix ${\bf B}$ given by 
\bea
{\bf B}=\left(\begin{array}{cc} d^{+}u_1^{+} & d^{-}u_1^{-}\nonumber \\
d^{+}u_2^{+} & d^{-}u_2^{-}\end{array}\right).
\eea
Therefore we have also that 
\bea
{\bf v}_i={\bf B}^{-1}{\bf k}_i
\label{lin}
\eea
with 
\bea
{\bf B}^{-1}=\frac{1}{m}\left(\begin{array}{cc} \frac{u_2^{-}}{u_2^{-}-u_1^{-}} & -\frac{u_1^{-}}{u_2^{-}-u_1^{-}}\nonumber \\ -\frac{u_2^{+}}{u_1^{+}-u_2^{+}} &\frac{u_1^{+}}{u_1^{+}-u_2^{+}}\end{array}\right)
\eea
which is always well defined except for values in parameter space of zero Lebesgue measure where $u^{+}_1=1$. 
Since $\lambda_+$ and $\lambda_-$have respectively positive and negative sign,  Eq.~$(\ref{lin})$ defines the two linear combination of the degrees $k^{[1]}$ and $k^{[2]}$ that respectively increases and decreases as a power-law of time.
Therefore we have found the solution of the model, once the constants $c_1$ and $c_2$ are given.
In order to find the correct values of the constants $c_1$ and $c_2$ given by Eqs.$(\ref{selfc0})$, we need to close our self-consistent argument.
Since we have assumed that the constants $c_1$ and $c_2$ are independent on the network realization, determined in the mean-field approximation by the quenched disorder of the assignment of the energies to the nodes,  the constants $c_1$ and $c_2$ can be  evaluated performing the following limits:

\bea
\lim_{t\to \infty}\frac{\Avg{\sum_j e^{-\beta\epsilon_j}k_j^{[1]}}}{mt}&=&c_1, \nonumber \\
\lim_{t\to \infty}\frac{\Avg{\sum_j e^{-\beta\epsilon_j}k_j^{[2]}}}{mt}&=&c_2,
\label{selfc1}
\eea
where in Eqs.~$(\ref{selfc1})$ the average is performed over the distribution of the energies of the nodes.
Therefore, by  multiplying each equation of Eq. $(\ref{k2})$ by $e^{-\beta\epsilon_i}$, integrating over the continuous time $t_i$, and averaging over the $g(\epsilon)$ distribution we get that  the self consistent equations determining the constants $c_1$ and $c_2$, or equivalently $\lambda_+$ and $\lambda_-$, are given by 
\bea
m{\bf c}={\bf B}{\bf J}
\label{dir}
\eea
where the column vector ${\bf J}=(J_+,J_-)$ has component given by 
\bea
J_+&=&\int d\epsilon g(\epsilon)\frac{1/\lambda_{+}}{e^{\beta\epsilon}/\lambda_{+}-1},\nonumber \\
J_-&=&\int d\epsilon g(\epsilon)\frac{1/\lambda_{-}}{e^{\beta\epsilon}/\lambda_{-}-1}.
\eea
Inverting  Eqs.~$(\ref{dir})$ we can express ${\bf J}$ as
\bea
{\bf J}&=&m{\bf B}^{-1}{\bf c}=\left(\begin{array}{c}\frac{(c_1u_2^{-}-c_2u_1^{-})}{u_2^{-}-u_1^{-}}\nonumber\\ \frac{(c_2u_1^{+}-c_1u_2^{+})}{u_1^{+}-u_2^{+}}\\\end{array}\right).
\eea
By defining the two constants $\mu_B$ and $\mu_F$, as in the following, 
\bea
\lambda^{+}=e^{\beta\mu_B}\nonumber \\
-\lambda^{-}=e^{\beta\mu_F}
\eea
and multiplying $J^{+}$ by $\lambda^{+}$ and $J^{-}$ by $-\lambda^{-}$ we get the following self-consistent equation, fixing the ``chemical potentials" $\mu_B$ and $\mu_F$,
\bea
I_B=\int d\epsilon g(\epsilon)\frac{1}{e^{\beta(\epsilon-\mu_B)}-1}=G_{B},\nonumber \\
I_F=\int d\epsilon g(\epsilon)\frac{1}{e^{\beta(\epsilon-\mu_F)}+1}=G_{F},
\label{selfq}
\eea
with  $G_B$ and $G_F$ independent on the energy distribution and only function of the inverse temperature $\beta$ and the two ``chemical potentials" $\mu_B$ and $\mu_F$. In fact we have,
\bea
{\bf G}&=&\left(\begin{array}{c}G_B\nonumber\\ G_F\end{array}\right)=\left(\begin{array}{c}\frac{(c_1u_2^{-}-c_2u_1^{-})}{u_2^{-}-u_1^{-}} \lambda_+\nonumber\\ \frac{(c_1u_2^{+}-c_2u_1^{+})}{u_1^{+}-u_2^{+}}\lambda_-\\\end{array}\right).
\eea
From the self-consistent Eq. $(\ref{selfq})$ the two constants $\mu_B$ and $\mu_F$ can be interpreted as `chemical potentials" of coupled Bose and Fermi gases and fully determine the evolution of the supersymmetric multiplex network, as long as the equations can be satisfied.
Only the left hand side of the Eqs. $(\ref{selfq})$  depends on the energy distribution $g(\epsilon)$ while the right hand side does not depend on it. Moreover the quantities   $G_B$ and $G_F$  depend on both the chemical potential $\mu_B$ and $\mu_F$ and can be explicitly expressed as 
\bea
G_B&=&e^{\beta\mu_B}\frac{(2-\alpha)e^{-\beta\mu_B}+\alpha e^{-\beta\mu_F}+(1-2\alpha)\sqrt{\Lambda}}{\alpha e^{-\beta\mu_B}+(2-\alpha)e^{-\beta\mu_F}+(1-2\alpha)\sqrt{\Lambda}}\times\nonumber\\
&&\times\frac{(1-\alpha)^2+\alpha^2}{2\alpha}\left[-e^{-\beta\mu_B}+e^{-\beta\mu_F}+\sqrt{\Lambda}\right],\nonumber\\
G_B&=&e^{\beta\mu_F}\frac{-\alpha e^{-\beta\mu_B}+(2-\alpha) e^{-\beta\mu_F}+(1-2\alpha)\sqrt{\Lambda}}{-(2-\alpha) e^{-\beta\mu_B}+\alpha e^{-\beta\mu_F}+(1-2\alpha)\sqrt{\Lambda}}\times\nonumber\\
&&\times\frac{(1-\alpha)^2+\alpha^2}{2\alpha}\left[-e^{-\beta\mu_B}+e^{-\beta\mu_F}+\sqrt{\Lambda}\right],\nonumber
\eea
where $\Lambda$ is given by 
\bea
\hspace*{-5mm}\Lambda=\left(e^{-\beta\mu_B}-e^{-\beta\mu_F}\right)^2+\frac{4\alpha^2}{[(1-\alpha)^2+\alpha^2]}e^{-\beta(\mu_B+\mu_F)}.
\eea
The self-consistent Eqs. $(\ref{selfq})$ that fix the chemical potential $\mu_B$ and $\mu_F$ fully determine the mean-field solution of this model.
In the supersymmetric multiplex network, nevertheless there can be two phenomena that implies a breakdown of this solution.
On one side we can observe a condensation of the links in correspondence of the regime of high values of $\beta$ where the Eqs.$(\ref{selfq})$ do not have a solution. This phenomenon will be discussed more in depth in the next section. On the other side, it is possible to observe in the model stochastic effects that are not captured by the mean-field solution.
\section{Structural properties of the supersymmetric multiplex network}
\label{sim}
The mean-field solution of the model well capture the main characteristics of the supersymmetric multiplex as long as $\alpha$ is not too large.
In fact we found very good agreement of the mean-field theory with the simulation results as long as $\alpha$ is lower than $0.5$.
For higher values of $\alpha$ in the second layer the rewiring process has a higher rate of the process of addition of new links and therefore non-trivial stochastic effects set in that are not captured by the mean-field solution.
For this reason, here we focus on the regime $\alpha<0.5$, where we find very good agreement between the theory and the simulations results.
We will describe the structural properties of the supersymmetric multiplex networks, covering the degree distribution of the networks in the two layers, different types of correlations  typical of multiplex networks, and we will describe the phenomenology related to the supersymmetric multiplex condensation transition in which one node acquires a finite fraction of all the links in both layers.
\subsection{ Degree distribution}
\begin{figure}
\begin{center}
\centerline{\includegraphics[width=3in]{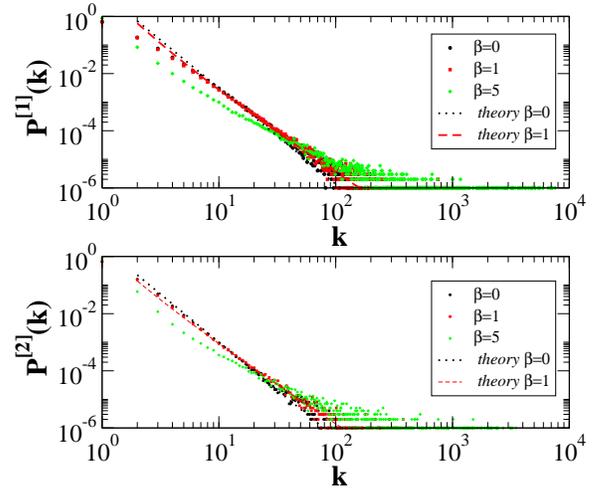}    }
\end{center}
\caption{The degree distributions $P^{[1]}(k)$ and $P^{[2]}(k)$ in the two layers of the supersymmetric network. The data is shown for  networks of $N=10^4$ nodes with energy distribution $g(\epsilon)=(1+\theta)\epsilon^{\theta}$ and $\theta=0.5$,  $\alpha=0.3$, $\beta=0,1,5$. The curves are averaged over $100$ multiplex networks realizations.The lines indicate the mean-field expectation for $\beta=0$ and $\beta=1$ and are in very good agreement with the simulation results.  The degree distribution for  $\beta=5$  is a typical degree distribution below the condensation transition where very big hubs emerge in the network. }
\label{fig_dd}
\end{figure}

The degree distribution in the network is scale-free in both layers, as predicted by the mean-field solution.
In fact if we consider the dynamical Eq. ($\ref{k1}$) for the degree $k^{[1]}_i$ and the degree $k^{[2]}_i$ and we take only the leading term in the limit $t\gg 1$ we found
\bea
k^{[1]}_i\simeq d^+u_1^{+}\left(\frac{t}{t_i}\right)^{e^{-\beta(\epsilon-\mu_B)}},\nonumber \\
k^{[2]}_i\simeq d^+u_2^{+}\left(\frac{t}{t_i}\right)^{e^{-\beta(\epsilon-\mu_B)}}.
\label{k12}
\eea 
Therefore, using the same mean-field arguments that are used to show that growing complex networks with preferential attachment are scale-free \cite{BA,Bose}, we can approximate the degree distributions $P^{[1]}(k)$ and $P^{[2]}(k)$ in the two layers as 
\bea
P^{[1]}(k)\propto\int d\epsilon g(\epsilon) e^{\beta(\epsilon-\mu_B)}\left(\frac{d^+u_1^+}{k}\right)^{e^{\beta (\epsilon-\mu_B)}}\frac{1}{k},\nonumber \\
P^{[2]}(k)\propto\int d\epsilon g(\epsilon) e^{\beta(\epsilon-\mu_B)}\left(\frac{d^+u_2^+}{k}\right)^{e^{\beta (\epsilon-\mu_B)}}\frac{1}{k}.
\eea
These expression reveals that the degree distributions in the two layers can be seen as a convolution of power-law networks with exponents $\gamma(\epsilon)=e^{\beta (\epsilon-\mu_B)+1}>2$.
In Figure $\ref{fig_dd}$ we show the degree distribution of the two layers for $g(\epsilon)=(\theta+1)\epsilon^{\theta}$ and $\theta=0.5$ for different values of $\beta=0.,1,5$.
The mean-field theory valid as long as the supersymmetric multiplex is not condensed, is in very good agreement with the simulation results.
 
\subsection{Multilayer degree correlations}
\begin{figure}
\begin{center}
\centerline{\includegraphics[width=3.3in]{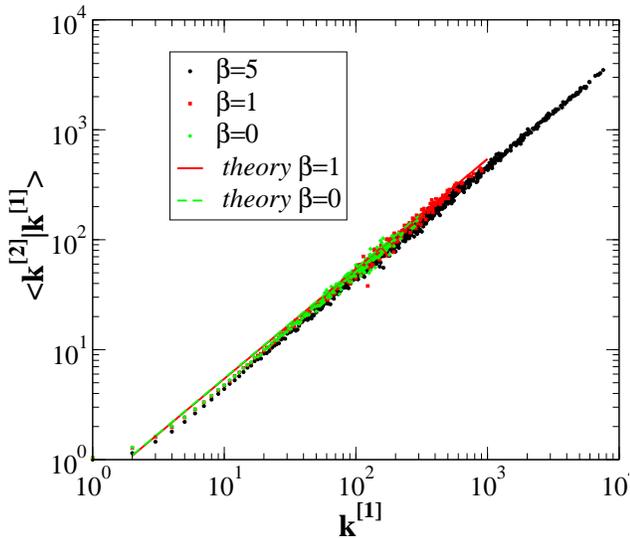}    }
\end{center}
\caption{The average degree in layer 2 conditioned on the degree in layer 1,  $\Avg{k^{[2]}|k^{[1]}}$ is shown and compared with the mean-field theoretical expectations finding very good agreement.  The data is shown for  networks of $N=10^4$ nodes, energy distribution $g(\epsilon)=(1+\theta)\epsilon^{\theta}$ with $\theta=0.5$,  $\alpha=0.3$ and $\beta=0,1,5$. The curves are averaged over $100$ multiplex networks realizations.}
\label{fig_avgk}
\end{figure}

In the multiplex networks one relevant correlation is between the degrees of the replica nodes.
In particular in a duplex  it is interesting to investigate if a hub in a network is also typically a hub in the other network or if it is typically a low degree node.
In the supersymmetric multiplex network model, we observe that for  $\alpha<0.5$, these correlations are positive.
In fact  in the mean-field solution, approximating the degrees  $k^{[1]}_i$ and $k^{[2]}_i$   for $t\to \infty$ as in Eq.~$(\ref{k12})$, we have 
\bea
k^{[2]}_i=\frac{u_2^{+}}{u_1^{+}}k^{[1]}_i,
\label{k2k1c}
\eea
i.e. the degree of layer 2 is positively and linearly correlated with the degree in layer 1.
In order to compare this mean-field expectations with the simulation results, we measure from the simulations results the average degree in layer 2 conditioned on the degree in layer 1, i.e. $\Avg{k^{[2]}|k^{[1]}}$.
This quantity characterizes the degree correlation in the multiplex network and is defined as 
\bea
\Avg{k^{[2]}|k^{[1]}}=\sum_{k^{[2]}}k^{[2]}P(k^{[2]}|k^{[1]}),
\eea
where $P(k^{[2]}|k^{[1]})$ is the conditional distribution of having a node of degree $k^{[2]}$ in layer 2 given that it has degree $k^{[1]}$ in layer 1.
Since in the mean-field approximation the degrees of  nodes are deterministic variables, the mean-field expectation for   $\Avg{k^{[2]}|k^{[1]}}$ is given by Eq~$(\ref{k2k1c})$.
In Figure $\ref{fig_avgk}$ we display   $\Avg{k^{[2]}|k^{[1]}}$ showing that is  an increasing function of $k^{[1]}$ indicating that the degrees of the same node  in the two layer of the supersymmetric multiplex are positively correlated as long as $\alpha<0.5$. Moreover, the conditional average  $\Avg{k^{[2]}|k^{[1]}}$ is well approximated by the mean-field expectation given by Eq.~$(\ref{k2k1c})$.
\subsection{ Bose-Einstein condensation in the supersymmetric multiplex network}
The Bianconi-Barab\'asi model \cite{Bose} describing a growing scale-free networks that can mapped to a Bose-Einstein gas, displays the Bose-Einstein condensation in complex networks. This condensation transition is a structural phase transition occurring in the network when the mapped Bose gas is in the the Bose-Einstein condensation  phase. Below this phase transition, in the network, one node grabs a finite fraction of the links and non trivial non-equilibrium process determine the network evolution. 

A similar phenomenon occurs also in the supersymmetric multiplex model, where  the condensation occurs simultaneously on the two replicas of the same node. In this model the condensation phase transition occurs at $\beta=\beta_c$ for which $\mu_B=0$. Therefore the equations determining the condensation phase transition are 
\bea
\left\{\begin{array}{l}I_B=\int d\epsilon g(\epsilon)\frac{1}{e^{\beta_c \epsilon}-1}=G^{B},\nonumber \\
I_F=\int d\epsilon g(\epsilon)\frac{1}{e^{\beta_c(\epsilon-\mu_F)}+1}=G^{F},\nonumber \\
\mu_B=0.\end{array}\right.
\eea   
Below this phase transition a single node grabs a finite fraction of all the links in both layers.
This condensation can possibly occur at low enough temperatures $T=1/\beta$ only if the integral $I_B(\mu_B=0)$ converges. Therefore, as in the classical Bose-Einstein condensation a necessary condition for this condensation to occur is that $g(\epsilon)\to 0$ as $\epsilon\to 0$.   

\begin{figure}
\begin{center}
\centerline{\includegraphics[width=4.1in]{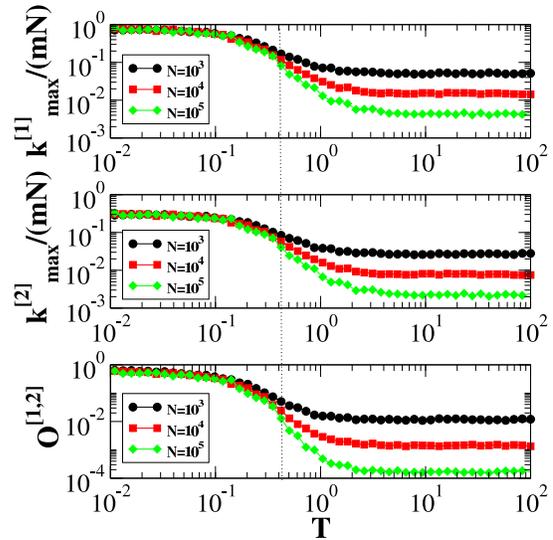}    }
\end{center}
\caption{The fraction of links $k^{[1]}_{max}/(mN)$ and $k^{[2]}_{max}/(mN)$ linked to the most connected node in layer 1 and in layer 2 are shown together with the total overlap of the links ${\cal O}^{[1,2]}$ versus $T=1/\beta$.The data is shown for  networks of $N=10^3,10^4,10^5$ nodes with  energy distribution $g(\epsilon)=(1+\theta)\epsilon^{\theta}$ and $\theta=0.5$,   $\alpha=0.3$. The curves are averaged over $100$ multiplex networks realizations for $N=10^3$ and $N=10^4$, and over $30$ multiplex network realization for $N=10^5$.}
\label{fig_cond}
\end{figure}

Since the condensation occurs in the same node in the two layers, we observe that below the condensation transition the two layers develop another type of correlation. In fact we observe that the total overlap of the links in the two layers becomes significant below the condensation transition. The total overlap ${\cal O}^{[1,2]}$ of the links between two layers (in this case layer 1 and layer 2)\cite{PRE,Note} is defined as 
\bea
{\cal O}^{[1,2]}=\sum_{i<j}a_{ij}^{[1]}a_{ij}^{[2]},
\eea
where $a_{ij}^{[1]}$ and $a_{ij}^{[2]}$ are the matrix elements of the adjacency matrix of layer 1 and layer 2 respectively.
In Figure~$\ref{fig_cond}$ we plot the fraction of the links linked to the most connected node in layer 1 and in layer 2. The absence of finite size effects below the condensation phase transition shows that in the supersymmetric multiplex there is one node that grabs a finite fraction of all the links. Moreover, in Figure~$\ref{fig_cond}$ we plot also  the total overlap ${\cal O}^{[1,2]}$of the links, showing that below the condensation transition the total overlap becomes significant.
\section{ Entropy rate of the supersymmetric multiplex network}
Given the initial condition, the supersymmetric multiplex evolution up to time $t$ is fully determined by the sequence of symbols $X=\{ \sigma_l^{[1]},i^{[1]}_l,\sigma^{[2]}_l,i_l^{[2]}\}_{l=1,2,\ldots mt}$.
Therefore, similarly to what happens for growing network models \cite{Entropyrate}, it is possible to define an entropy rate of the growing supersymmetric multiplex network.
This entropy rate $H(X)$ can be useful for example if aim at compressing the network, as we could aim at extending  the Shannon's noiseless coding  theorem \cite{Quantuminformation}  to  the sequence $X=\{ \sigma_l^{[1]},i^{[1]}_l,\sigma^{[2]}_l,i_l^{[2]}\}_{l=1,2,\ldots mt}$ encoding the full network evolution.
The entropy rate of the supersymmetric multiplex network model when in the supersymmetric multiplex we have already observed $l-1$ network changes,  is given by 
\bea
H(X)&=&-\sum_{{\bf x}_l}P({\bf x}_l|\{{\bf x}_{\ell}\}_{\ell=1,\ldots, l-1})\nonumber \\
&&\times\log P({\bf x}_l|\{{\bf x}_{\ell}\}_{\ell=1,\ldots, l-1}),
\label{er}
\eea
where ${\bf x}_l=\{ \sigma_l^{[1]},i^{[1]}_l,\sigma^{[2]}_l,i_l^{[2]}\}$.
We note here that the entropy rate of the supersymmetric multiplex, as the entropy rate of growing networks has a very characteristic feature, i.e. contains a leading term of order of $\log (t)$. Therefore it does not converge in the thermodynamic limit $t\to \infty$.
This is due to the fact that the attachment probability in these networks are non-local.
This occurs also in the growth of other networks \cite{Entropyrate} such as random trees, where each new node is attached to a random node of the network with probability $1/t$, where $t$ is the number of  nodes in the network.
In fact it is easy to see that also for this basic, non-local model we have 
\bea
H(X)=\log (t).
\eea

\section{ The entanglement entropies of the layers  of the supersymmetric multiplex network }
In order to evaluate the complexity of a single layer, recently new attention has been devoted to quantum information measures \cite{Vonneuman,Garnerone,Jesus2}. Already  the von Neumann entropy \cite{Vonneuman} of single networks has been extended to multilayer networks in Ref. \cite{Math}.
Here we propose to consider the entanglement entropies of the layers of the supersymmetric multiplex network as a generalization of the quantum entropy proposed in   <ref. \cite{Garnerone}.
Following \cite{Garnerone} here we  perform a  mapping between the layers of the supersymmetric multiplex   network with $M=2$ and two  bipartite quantum states.
In particular we will consider the states $\ket{A^{[1]}},\ket{A^{[2]}} \in {\cal H}_{a} \otimes {\cal H}_b$ with ${\cal H}_a \simeq {\cal H}_b \simeq \mathbb{C}^N$, given by 
\bea
\ket{A^{[1]}}&=&\frac{1}{\parallel A^{[1]}\parallel}_F\sum_{i,j} e^{-\beta \epsilon_i/2} a^{[1]}_{ij}\ket{i}\ket{j}\nonumber\\
\ket{A^{[2]}}&=&\frac{1}{\parallel A^{[2]}\parallel}_F\sum_{i,j} e^{-\beta \epsilon_i/2} a^{[2]}_{ij}\ket{i}\ket{j}
\eea
where ${\bf a^{[1]}}$ is the adjacency matrix of the first layer of the supersymmetric multiplex network,  ${\bf a^{[2]}}$ is the adjacency matrix of the second layer of the supersymmetric multiplex network. Moreover,  $\parallel A\parallel=\sqrt{\mbox{Tr}{\bf A}^{\dag}{\bf A}}$ denotes the Frobenious norm of the matrix ${\bf A}$ and the matrix elements $(i,j)$ of the matrices ${\bf A^{[1]}}$ and ${\bf A^{[2]}}$ are given respectively by $A^{[1]}_{ij}=e^{-\beta\epsilon_i/2}a^{[1]}_{ij}$ and $A^{[2]}_{ij}=e^{-\beta\epsilon_i/2}a^{[2]}_{ij}$.
Using the terminology of \cite{Garnerone} we will refer to $\ket{A^{[1]}}$ and $\ket{A^{[2]}}$ as pure network states.

In order to  characterize the complexity of the layers we propose to evaluate the entanglement entropy of the pure network states  $\ket{A^{[1]}}$ and $\ket{A^{[2]}}$. Therefore we define the 
reduced density matrices $\rho^{[1]},\rho^{[2]}$ given by 
\bea
\rho^{[1]}&=&\mbox{Tr}_b \ket{A^{[1]}}\bra{A^{[1]}}, \nonumber \\
\rho^{[2]}&=&\mbox{Tr}_b \ket{A^{[2]}}\bra{A^{[2]}}
\eea
and we calculate the entanglement entropy $S^{[1]}$ and $S^{[2]}$  given by 
\bea
S^{[1]}&=&-\mbox{Tr}_a \rho^{[1]}\log \rho^{[1]},\nonumber \\
S^{[2]}&=&-\mbox{Tr}_a \rho^{[2]}\log \rho^{[2]}.
\eea
Using the explicit expression for the reduced density matrices in terms of the degree of the nodes in the different layers $\rho^{[1]}$ and $\rho^{[2]}$, i.e.
\bea
\rho^{[1]}&=&\sum_i \frac{e^{-\beta \epsilon_i}k_i^{[1]}}{\sum_j e^{-\beta \epsilon_j}k_j^{[1]}}\ket{i}\bra{i}\nonumber \\
\rho^{[2]}&=&\sum_i \frac{e^{-\beta \epsilon_i}k_i^{[2]}}{\sum_j e^{-\beta \epsilon_j}k_j^{[2]}}\ket{i}\bra{i}
\eea
we found that the entropies $S^{[1]}$ and $S^{[2]}$ are given by 
\bea
S^{[1]}&=&-\sum_i \frac{e^{-\beta \epsilon_i}k_i^{[1]}}{\sum_j e^{-\beta \epsilon_j}k_j^{[1]}}\log\left( \frac{e^{-\beta \epsilon_i}k_i^{[1]}}{\sum_j e^{-\beta \epsilon_j}k_j^{[1]}}\right),\nonumber \\
S^{[2]}&=&-\sum_i \frac{e^{-\beta \epsilon_i}k_i^{[2]}}{\sum_j e^{-\beta \epsilon_j}k_j^{[2]}}\log\left( \frac{e^{-\beta \epsilon_i}k_i^{[2]}}{\sum_j e^{-\beta \epsilon_j}k_j^{[2]}}\right).
\eea
Moreover, in the asymptotic limit $t\gg 1$ we can evaluate the entropy using the asymptotic relations given by Eqs.~$(\ref{asymptotics})$.
We found that 
\bea
E^{[1]}-TS^{[1]}+T\log (mc_1t)=F^{[1]}\nonumber \\
E^{[2]}-TS^{[2]}+T\log (mc_2t)=F^{[2]}
\eea
where 
\bea
E^{[\nu]}&=&\avg{\epsilon}_{\nu}\nonumber \\
F^{[\nu]}&=&-T\avg{\log k^{[\nu]}}_{\nu}
\eea
with $\nu=1,2$ and
\bea
\Avg{f}_{\nu}=\sum_i  \frac{e^{-\beta \epsilon_i}k_i^{[\nu]}}{\sum_j e^{-\beta \epsilon_j}k_j^{[\nu]}} f_i
\eea
Finally by using the mean-field solution of the model we can evaluate the energies $E^{[1]},E^{[2]}$, and the free energies $F^{[1]},F^{[2]}$, as long as the supersymmetric multiplex is not in the condensed phase and $\alpha<0.5$.
If we define $E_B$ and $E_F$ respectively as the average energies calculated over the Bose and Fermi distributions with ``chemical potentials" $\mu_B$ and $\mu_F$, i.e.
\bea
E_B=\int d\epsilon g(\epsilon)
\frac{\epsilon}{e^{\beta(\epsilon-\mu_B)}-1}\nonumber \\
E_F=\int d\epsilon g(\epsilon)
\frac{\epsilon}{e^{\beta(\epsilon-\mu_F)}+1}
\eea we have, in the mean-field approximation, 
\bea
\left(\begin{array}{c}E^{[1]}\nonumber \\E^{[2]}\end{array}\right)={\bf V}\left(\begin{array}{c}E_B\nonumber \\E_F\end{array}\right).
\eea
with 
\bea
{\bf V}&=&\left(\begin{array}{cc}\frac{1}{mc_1}d^{+}u_1^{+}\lambda^{+}& -\frac{1}{mc_1}d^{-}u_1^{-}\lambda^{-}\nonumber \\ \frac{1}{m c_2}d^{+}u_2^{+}\lambda^{+}& -\frac{1}{mc_2}d^{-}u_2^{-}\lambda^{-} \end{array}\right).
\eea
Similarly also   the ``free energies" $F^{[1]}$ and $F^{[2]}$ can be estimated using the mean-field solution of the model. 
\section{Relation between the entanglement entropies of the supersymmetric multiplex and its entropy rate}
We note here a surprising result. In fact the entanglement entropies $S^{[1]}$ and $S^{[2]}$ have a immediate classical meaning because they can be linked to the entropy rate of the network evolution.
In fact, by calculating explicitly the entropy rate of the supersymmetric multiplex, defined in Eq. $(\ref{er})$, we get 
\bea
H(X)=S^{[1]}+S^{[2]}+2h(\alpha)
\eea
where  $h(\alpha)$ is given by 
\begin{equation}
h(\alpha)=-\alpha\log (\alpha)-(1-\alpha)\log (1-\alpha).
\end{equation}
Therefore, the entanglement entropies  of the supersymmetric multiplex network  are related to the entropy rate of the supersymmetric multiplex,  which is described by a classical  non-equilibrium process.
The relation between the entropy rate $H(X)$ and the entanglement entropies $S^{[1]}$ and $S^{[2]}$ remains valid for every value of $\alpha\in[0,1)$ and also below the condensation phase transition. Nevertheless, the scaling of the entanglement entropies with the system size changes below the condensation phase transition.
 
\section{ Conclusions}
In conclusion here we have  investigated the properties of the supersymmetric multiplex network model in which nodes have intrinsic fitness and the evolution describes both the addition of new links according to the generalized preferential attachment, and rewring of the links.
The resulting multiplex network has scale-free layers and develops interesting degree-degree correlations. 
The supersymmetric multiplex model  can be fully characterized by coupled quantum Bose-Einstein and Fermi-Dirac statistics.
In fact the dynamic rules of the supersymmetric multiplex networks evolution exploit the symmetries of the multilayer structure and, as a consequence of this, the multiplex network evolution is not determined exclusively by the Bose-Einstein statistics or by the Fermi-Dirac statistics, but is determined by  Bose-Einstein and Fermi-Dirac statistics with  coupled ``chemical potentials" $\mu_B$ and $\mu_F$.
The resulting supersymmetric multiplex network  can undergo a Bose-Einstein condensation of the links in which one node acquires a finite fraction of the links in all the layers, and simultaneously every pair of layers develops a significant overlap of the links.
Moreover, an interesting relation  has been shown to exists between the entanglement entropies of the layers in the supersymmetric multiplex network, measuring the complexity of these layers with quantum information theory tools, and the entropy rate of the classical supersymmetric multiplex network.
In conclusion, in this work the evolution of supersymmetric multiplex networks with fitness of the nodes is characterized. The   complexity of the supersymmetric multiplex networks, and its underlying symmetries have been shown to be related to quantum statistics. In fact  in multilayer networks there are additional symmetries  that are not present in undirected single networks. These symmetries   allow for an evolution determined simultaneously by Bose-Einstein and Fermi-Dirac statistics. Moreover, interesting results relate the complexity of these structures measured by quantum information theory tools and their non equilibrium classical dynamics determined by their entropy rate.

\appendix
\section{Supersymmetric multiplex networks with $M$ layers}
We consider here the extension of the supersymmetric multiplex model defined in the main text for a multiplex network of $M=2$ layers to the case in which the multiplex is formed by a generic number $M$ of layers.
We suppose that the layers can be distinguished in two groups: a first group of $M_1$ layers $\xi=1,2\ldots, M_1$ in which the dynamics only include addition of new links, a second group of $M_2$ layers $\psi=M_1+1,M_1+2,\ldots, M_1+M_2$ in which the network dynamics includes both addition of new links are rewiring of the links.
Clearly we must have $M_1+M_2=M$.
In particular we consider the following model.
We   start at $t=0$ from a small set of nodes $N_0$ connected in  each of the $M$ layers. 
Each node $i$ of the network has degrees $k^{[\xi]}_i$,  in the layers $\xi=1,2,\ldots, M_1$ and degrees $k^{[\psi]}_i$ in the layers $\psi=M_1+1,M_1+2\ldots, M_1+M_2$. Moreover each node has  an energy $\epsilon_i$ drawn from a $g(\epsilon)$ distribution and a fitness given by $\eta_i=e^{-\beta\epsilon_i}$.
At each time  $t$ we add a node  to the multiplex network, each node has $M$ replicas nodes, one on each layer and each replica node is  initially attached to $m$ existing nodes in the same layer. In the following we will  indicate with $l$ the number of network changes, i.e. links additions or link rewirings,  occurring in each layer  starting from time $t=1$. After time $t$ we will have $l=mt$.
For each   network change in a layer $\xi$  we follow the subsequent procedure:
\begin{itemize}
\item We extract a number $\sigma^{[\xi]}_l=\phi$ with $\phi=1,2,\ldots, M$. The event $\sigma^{[1]}_l=\xi'$ with $\xi'=1,2,\ldots, M_1$ occurs  with  probability  $p^{[\xi]}(\xi')=\alpha/M_1$ while the event $\sigma_l^{[\xi]}=\psi'$ with $\psi'=M_1+1,\ldots, M_1+M_2$ occurs with probability   $p^{[\xi]}(\psi')=(1-\alpha)/M_2$.
\item If $\sigma^{[\xi]}_l=\xi'$, the new node is attached  in layer $\xi$  to a node $i^{[\xi]}_l$ chosen with probability
\bea
\Pi^{[\xi,\xi']}\left(i_l^{[\xi]}=i\right)=\frac{e^{-\beta \epsilon_i}k_i^{[\xi']}}{\sum_j e^{-\beta \epsilon_j}k_j^{[\xi']}},
\eea
i.e. it will be attached preferentially to nodes with low energy and high degree in layer $\xi'$,  according to a generalized preferential attachment driven by layer $\xi'$.
Instead, if $\sigma^{[1]}_l=\psi'$ the new  node  is attached in layer $\xi$ to a node $i^{[\xi]}_l$ chosen with probability
 \bea
\Pi^{[\xi,\psi']}\left(i_l^{[\xi]}=i\right)=\frac{e^{-\beta \epsilon_i}k_i^{[\psi']}}{\sum_j e^{-\beta \epsilon_j}k_j^{[\psi']}}.
\eea
In other words the new node  will be attached preferentially to nodes with low energy and high degree in layer $\psi'$,  according to  a generalized preferential attachment.
\end{itemize}
For each  network change in layer $\psi$ we follow the subsequent procedure:
\begin{itemize}
\item  We extract a number $\sigma^{[\psi]}_l=\phi$ with $\phi=1,2\ldots, M$.
 The event $\sigma^{[\psi]}_l=\xi'$ with $\xi'=1,2,\ldots, M_1$ occurs  with  probability  $p^{[\psi]}(\xi')=(1-\alpha)/M_1$ while the event $\sigma_l^{[\psi]}=\psi'$ with $\psi=M_1+1,\ldots, M_1+M_2$ occurs with probability   $p^{[\psi]}(\psi')=\alpha/M_2$. 
\item If $\sigma^{[\psi]}_l=\xi'$  the new node will be attached, in  layer $\psi$, to a node $i_l^{[\psi]}$ chosen with probability
\bea
\Pi^{[\psi,\xi']}\left(i_l^{[\psi]}=i\right)=\frac{e^{-\beta \epsilon_i}k_i^{[\xi']}}{\sum_j e^{-\beta \epsilon_j}k_j^{[\xi']}}.
\eea
Instead, if $\sigma^{[\psi]}_l=\psi'$  a  random link of a node $i^{[\psi]}_l$ chosen with probability 
\bea
\Pi^{[\psi,\psi']}\left(i_l^{[\psi]}=i\right)=\frac{e^{-\beta \epsilon_i}k_i^{[\psi']}}{\sum_j e^{-\beta \epsilon_j}k_j^{[\psi']}}
\eea
 is rewired, i.e. it is detached from node $i_l^{[\psi]}$ and attached to the new node of the network.
\end{itemize}
Therefore the network is  determined by the sequence of the values $\{\sigma^{[\xi]}_l,i_l^{[\xi]}\sigma^{[\psi]}_l,i_l^{[\psi]}\}_{\xi=1,2\ldots, M_1, \psi'=M_1+1\ldots, M_1+M_2}$ that fully determines the evolution of the multiplex network given the initial condition.

The mean-field treatment of the model can be performed exactly has in the case of the supersymmetric multiplex networks formed by $M=2$ layers.
In fact the mean field equations for the degree in each layer read
\bea
\frac{dk_i^{[\xi]}}{dt}&=&me^{-\beta \epsilon_i}\left[\alpha\frac{1}{M_1}\sum_{\xi'}\frac{1}{\sum_j e^{-\beta\epsilon_j}k_j^{[\xi']}}k_i^{[\xi']}\right.\nonumber \\
&&\left.+(1-\alpha)\frac{1}{M_2}\sum_{\psi'}\frac{1}{\sum_j e^{-\beta\epsilon_j}k_j^{[\psi']}}k_i^{[\psi']}\right],\nonumber \\
\frac{dk_i^{[\psi]}}{dt}&=&me^{-\beta \epsilon_i}\left[\frac{(1-\alpha)}{M_1}\sum_{\xi'}\frac{1}{\sum_j e^{-\beta\epsilon_j}k_j^{[\xi']}}k_i^{[\xi']}\right.\nonumber \\
&&\left.-\frac{\alpha}{M_2}\sum_{\psi'}\frac{1}{\sum_j e^{-\beta\epsilon_j}k_j^{[\psi']}}k_i^{[\psi']}\right].
\label{mf0g}
\eea
We note that in the mean field equation the degree of the nodes in the layer $\xi=1,2\ldots, M_1$ are all the same, while the degree of the nodes in the layers $\psi=M_1+1,\ldots, M_1+M_2$ are also all the same, therefore we can write the mean-field equation for the average degree in the first group of layers $k_i^{[\xi]}=\kappa_i^{[1]} \ \forall \xi=1,2\ldots,M_1$ and the average degree in the second group of layers $k_i^{[\psi]}=\kappa_i^{[2]}\  \forall \psi=M_1+1,\ldots, M_1+M_2$.
We have in particular
\bea
\frac{d\kappa_i^{[1]}}{dt}=me^{-\beta \epsilon_i}\left[\frac{\alpha}{\sum_j e^{-\beta\epsilon_j}\kappa_j^{[1]}}\kappa_i^{[1]}+\frac{(1-\alpha)}{\sum_j e^{-\beta\epsilon_j}\kappa_j^{[2]}}\kappa_i^{[2]}\right],\nonumber \\
\frac{d\kappa_i^{[2]}}{dt}=me^{-\beta \epsilon_i}\left[\frac{(1-\alpha)}{\sum_j e^{-\beta\epsilon_j}\kappa_j^{[1]}}\kappa_i^{[1]}-\frac{\alpha}{\sum_j e^{-\beta\epsilon_j}\kappa_j^{[2]}}\kappa_i^{[2]}\right].
\label{mf0reduced}
\eea
These equations read completely equivalent to the mean-field equations Eqs. $(\ref{mf0})$ for the supersymmetric multiplex network of $M=2$ layers. 
Moreover it is straightforward to generalize the definition of the entanglement entropies of each layer, introduced in the main text for the $M=2$ case, for this general case.
It is immediate to see that also in this general case the sum of the entanglement entropies of each layer of the supersymmetric multiplex are linearly related with its entropy rate.
This last result is independent on the validity of the mean-field approach and is a fundamental characteristic of this model.

\begin{thebibliography}{99}
\bibitem{Note}
This definition is valid for unweighted layers, for a  generalization to weighted layers see \cite{Weighted}.
\end{thebibliography}
\end{document}